\documentclass[useAMS,usenatbib]{mnras}

\usepackage{epsfig}
\usepackage{aas_macros}
\usepackage{natbib}
\usepackage{times}
\usepackage{graphicx, bm, amssymb, xcolor}
\usepackage{verbatim}
\usepackage[all]{hypcap}
\usepackage{xspace}
\usepackage{multirow}
\usepackage{amsmath}
\usepackage{epstopdf}
\usepackage{pdflscape}
\usepackage{xspace}

\newcommand{\msun}{\mbox{M$_\odot$}}

\newcommand{\gyr}{\mbox{${\rm Gyr}$}}
\newcommand{\pc}{\mbox{${\rm pc}$}}

\newcommand{\be}{\begin{equation}}
\newcommand{\ee}{\end{equation}}
\newcommand{\bea}{\begin{eqnarray}}
\newcommand{\eea}{\end{eqnarray}}

\newcommand{\emosaics}{{\sc E-MOSAICS}\xspace}
\newcommand{\mosaics}{{\sc MOSAICS}\xspace}
\newcommand{\eagle}{{\sc EAGLE}\xspace}

\title{\vspace{-6mm}Dynamical cluster disruption and its implications for multiple population models in the E-MOSAICS simulations\vspace{-5mm}}
\author{M.~Reina-Campos$^{1}$\thanks{reina.campos@uni-heidelberg.de}, J.~M.~D.~Kruijssen$^{1}$, J.~Pfeffer$^{2}$, N.~Bastian$^{2}$ and R.~A.~Crain$^{2}$\\
$^{1}$Astronomisches Rechen-Institut, Zentrum f\"{u}r Astronomie der Universit\"{a}t Heidelberg, M\"{o}nchhofstra\ss e 12-14, 69120 Heidelberg, Germany\\
$^{2}$Astrophysics Research Institute, Liverpool John Moores University, 146 Brownlow Hill, Liverpool L3 5RF, UK}

\begin{document}

\date{\vspace{-6mm}}

\pagerange{\pageref{firstpage}--\pageref{lastpage}} \pubyear{2018}

\maketitle

\label{firstpage}

\begin{abstract}
Several models have been advanced to explain the multiple stellar populations observed in globular clusters (GCs). Most models necessitate a large initial population of unenriched stars that provide the pollution for an enriched population, and which are subsequently lost from the cluster. This scenario generally requires clusters to lose $>90$~per~cent of their birth mass. We use a suite of 25 cosmological zoom-in simulations of present-day Milky Way-mass galaxies from the \emosaics project to study whether dynamical disruption by evaporation and tidal shocking provides the necessary mass loss. We find that GCs with present-day masses $M>10^5~\msun$ were only $2$--$4$ times more massive at birth, in conflict with the requirements of the proposed models. This factor correlates weakly with metallicity, gas pressure at birth, or galactocentric radius, but increases towards lower GC masses. To reconcile our results with observational data, either an unphysically steep cluster mass-size relation must be assumed, or the initial enriched fractions must be similar to their present values. We provide the required relation between the initial enriched fraction and cluster mass. Dynamical cluster mass loss cannot reproduce the high observed enriched fractions nor their trend with cluster mass.  
\end{abstract}


\section{Introduction} \label{sec:intro}

Over the past decades, both photometric and spectroscopic studies have indicated chemical anomalies in the stellar populations of globular clusters (GCs). The presence of multiple main sequences in the optical colour magnitude diagrams of GCs \citep[e.g.][]{bedin04,piotto07,milone12,piotto15} suggest differences in helium content, whereas spectroscopic studies \citep[e.g.][]{carretta09a,gratton12,carretta15} have revealed anti-correlations in the chemical abundances of light-element species (Al, Na, N, C, Mg). Most models aiming to explain the prevalence of multiple populations in GCs assume the formation of a first generation of stars, which rapidly pollute the clusters residual gas, from which a second generation of stars is born. The mechanism through which the medium is enriched is a question of ongoing debate and depends on the model considered. Some authors suggest AGB stars are the polluters (e.g. \citealt{dercole08}), whereas others consider stellar winds from fast rotating massive stars (FRMS, e.g. \citealt{krause13}), supermassive stars (SMSs, \citealt{denissenkov14}) or massive binaries (\citealt{demink09}) to be the source of the medium enrichment. See \cite{bastian17} for a review.

These models encounter many observational challenges, but we will focus on two of them here. Firstly, the proposed polluting stars make up only a small fraction of a standard stellar population due to the stellar initial mass function (IMF).  However, the median observed fraction of enriched stars is $\sim 64$ per cent (\citealt{milone17}). Hence, in order to work, the models require that GCs were originally much more massive than today, and have lost $>90$--$95$ per cent of their initial masses. This is known as the \textit{mass budget} problem and we will refer to the affected models as mass budget-limited (MBL). Secondly, we expect the fraction of the GC mass lost to be strongly environmentally dependent. However, observed fractions of enriched stars only increase with the GC mass and show no or little correlation with other quantities \citep{carretta10b,conroy12,kruijssen15b,bastian15,milone17}.

In this paper, we study whether dynamical cluster disruption mechanisms provide the mass loss necessary to reproduce the observed enriched fractions and its variation with GC properties, and compare it to observational data. We also use our results to reverse-engineer what the initial enriched fraction must have been in order to reproduce the observed enriched fractions at the present day.

\begin{figure*}
\centering
\includegraphics[width=\hsize,keepaspectratio]{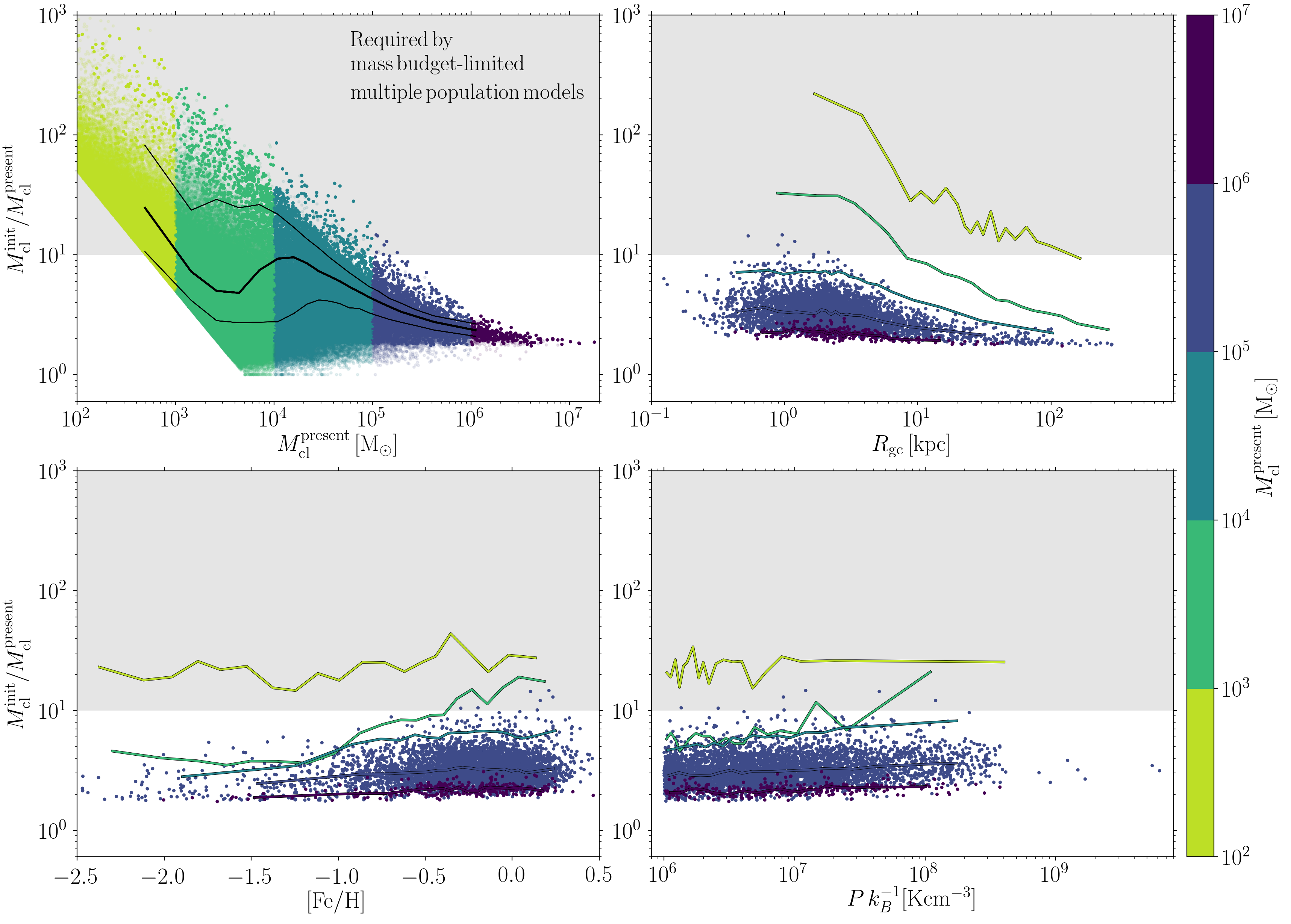}%
\vspace{-2mm}\caption{\label{fig:mass-loss-pressure} Initial-to-present cluster mass ratio as a function of present cluster mass (\textit{top left}), present galactocentric radius (\textit{top right}), cluster metallicity (\textit{bottom left}) and gas pressure at birth (\textit{bottom right}). All clusters with Milky Way-like metallicities ($\rm [Fe/H]\in[-2.5,0.5]$) that survive to the present time in our 25 haloes are included in the top left panel, but only the most massive clusters ($M>10^5~\msun$) are shown in the rest of panels. We emphasize the old ($\tau>6~\gyr$) clusters born in high-pressure environments ($P\,k_{\rm B}^{-1}>10^6\,\rm K\,cm^{-3}$) in all the panels with opaque data points, whereas the rest are transparent and only appear in the top left panel. The thick and thin solid lines in the top left panel correspond to the median and the $1\sigma$ dispersion of the opaque data, whereas the solid lines in the rest of panels represent the opaque data. The grey area indicates the mass loss required by MBL multiple population models for solving the mass budget problem.}\vspace{-3mm}
\end{figure*}

\section{Dependence of cluster mass loss on the galactic environment}

We probe the dependence of cluster mass loss on the galactic environment using the 25 cosmological zoom-in simulations from the \emosaics project which focuses on the evolution of present-day Milky Way-mass galaxies (defined by the halo mass range $11.86<\log(M_{200}/\msun)<12.38$, \citealt{pfeffer18}, \citealt{kruijssen18b}). The \emosaics project couples \mosaics (MOdelling Star cluster population Assembly In Cosmological Simulations, \citealt{kruijssen11}), a sub-grid model for stellar cluster formation and evolution, to the \eagle (Evolution and Assembly of GaLaxies and their Environments, e.g. \citealt{schaye15}, \citealt{crain15}) galaxy formation model. \eagle is a suite of hydrodynamical simulations of galaxy formation in the $\Lambda$CDM cosmogony, evolved using a modified version of the N-body TreePM smoothed particle hydrodynamics (SPH) code \textsc{GADGET-3} (\citealt{springel05c}). The suite of galaxies from the \emosaics project are the first simulations to describe self-consistently the formation and evolution of stellar clusters in a cosmological context. In this project, stellar clusters are formed as a sub-grid component of the stellar particles following an environmentally-dependent cluster formation efficiency (\citealt{kruijssen12d}) and initial cluster mass function (\citealt{reina-campos17}). The stellar clusters then lose mass due to stellar evolution (\citealt{wiersma09b}), tidal shocking and evaporation (\citealt{kruijssen11}). In addition, we apply destruction of the most massive GCs by dynamical friction in post-processing (see \citealt{pfeffer18} and \citealt{kruijssen18b} for further details of the simulations). The philosophy behind \emosaics is to describe the formation and evolution of GCs using the physical constraints obtained from observations of star formation, young massive clusters (YMCs), and GCs from low to high redshift.

Observations, theory, and simulations together show that the amount of cluster mass loss is expected to depend on several factors (see \citealt{kruijssen15b} for a detailed discussion). Firstly, we expect a strong dependence on the cluster mass, i.e.~massive clusters lose less mass than low mass clusters. Secondly, three other factors have significant, but weaker effects. We expect a dependence on the gas pressure at birth, because higher-pressure environments imply larger gas density contrasts \citep[e.g.][]{vazquez94,padoan17}, so stronger tidal interactions are expected that will generally lead to more efficient mass loss due to tidal shocks. We also expect a dependence on galactocentric radius, because the tidal field strengths and gas pressure both decrease outwards. Finally, we expect a dependence on metallicity, because it traces the host galaxy mass (albeit with large scatter) and more massive galaxies are generally characterised by stronger tidal fields and higher gas pressures. The emergence of these dependences is modelled self-consistently in \emosaics.

\begin{figure*}
\centering{}
\includegraphics[width=\hsize,keepaspectratio]{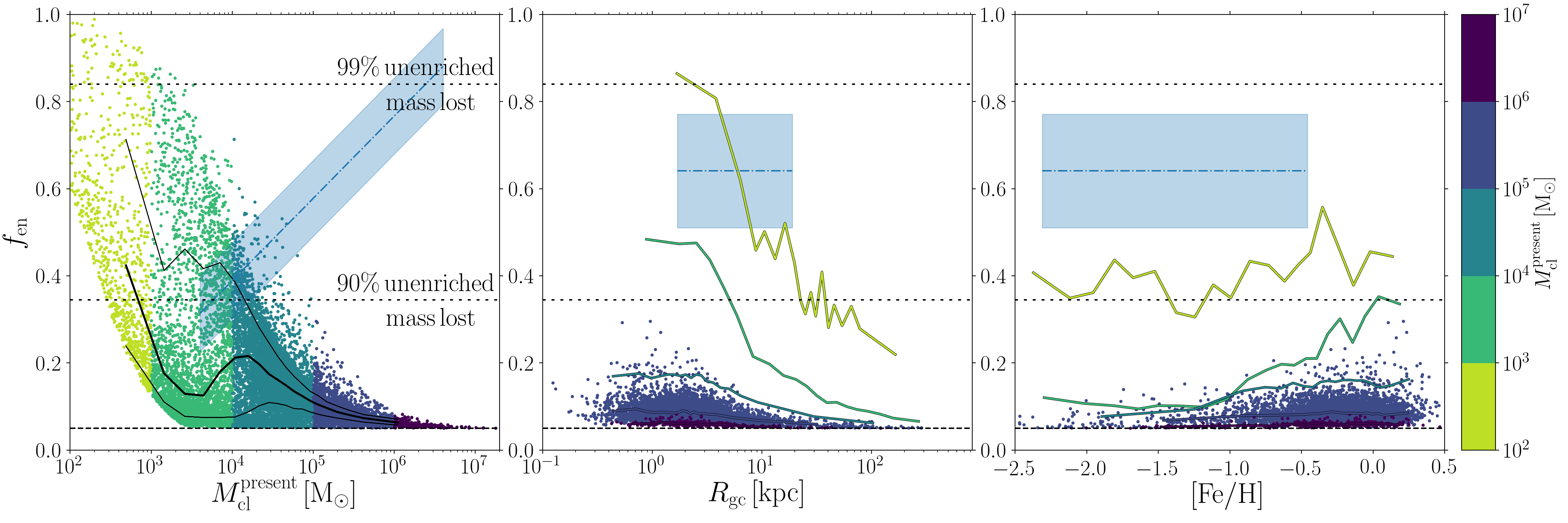}%
\vspace{-2mm}\caption{\label{fig:fenriched-pressure} Fraction of enriched stars (Eq.~\ref{eq:fenriched}) as a function of present cluster mass (\textit{left}), present galactocentric radius (\textit{middle}) and cluster metallicity (\textit{right}). Data points and lines have the same meaning as in Fig.~\ref{fig:mass-loss-pressure}. The blue dash-dotted line and blue shaded area in the left panel correspond to a fiducial observed relation and the $1\sigma$ dispersion around the fit obtained from \citet{milone17}, ${f_{\rm en}} = 0.189\log(M/{M_\odot})-0.367$, whereas the blue dash-dotted line and blue shaded area in the middle and right panels correspond to the median enriched fraction and the $1\sigma$ dispersion around the median for the same cluster sample, respectively. The black dashed line corresponds to the initial enriched fraction $f_{\rm en}^{\rm init}=5$ per cent typically assumed by MBL multiple population models. The upper and lower dotted black line corresponds to a 99 and 90 per cent mass loss, respectively, in the form of unenriched stars.}\vspace{-3mm}
\end{figure*}

The absence of an explicit model for the cold, dense phase of the interstellar medium in \eagle, which is predicted to dominate the disruptive power of galaxies, implies an underestimation of cluster disruption in our simulations. Though this could present a major caveat for their use to estimate the true mass loss of star clusters, \citet{pfeffer18} show that we obtain a reasonable estimate of disruption in high-pressure environments ($P\,k_{\rm B}^{-1}>10^6\,\rm K\,cm^{-3}$). Therefore, we focus our analysis on old ($\tau>6~\gyr$) GCs born in these high gas pressure environments, for which disruption is expected to be strongest and \emosaics provides a good approximation of the total amount of mass loss. 

We determine the initial-to-present mass ratio of all stellar clusters with Milky Way-like metallicities ($\rm [Fe/H]\in[-2.5,0.5]$) that survive to the present time in each of our 25 galaxies, and we show the results as a function of the present cluster mass in Fig.~\ref{fig:mass-loss-pressure}. At fixed cluster mass, we also determine the median mass ratio as a function of present galactocentric radius, cluster metallicity and gas pressure at birth to investigate secondary dependences. The hard cut at low masses in the top left panel results from the minimum initial cluster mass ($M_{\rm cl}\geq5\times10^3~\msun$) adopted in \emosaics for explicitly modelling the evolution of individual clusters.

Fig.~\ref{fig:mass-loss-pressure} shows a steeply decreasing trend of the initial-to-present mass ratio as a function of the cluster mass for clusters more massive than $\gtrsim10^4~\msun$, and weak trends in the expected direction with galactocentric radius, cluster metallicity and gas pressure at birth for intermediate mass ($10^3$--$10^5~\msun$) clusters. For the massive clusters ($M>10^5~\msun$), the lack of a trend between the median initial-to-present mass ratio and all cluster properties other than its mass arises because these clusters lose little mass, of the order of a factor $2$--$5$, as the relative mass loss rate is inversely proportional to the cluster mass for a constant radius. As a result, most of the mass in high-mass clusters is lost due to stellar evolution. MBL multiple population models require a minimum initial-to-present mass ratio of $M_{\rm cl}^{\rm init}/M_{\rm cl}^{\rm present} \ga 10$ to solve the mass budget problem. Given this constraint, only clusters with $M_{\rm cl}<10^5~\msun$ should exhibit the high observed enriched fractions \citep[also see][]{kruijssen15b}, whereas observed enriched fractions increase with cluster mass (e.g.~see fig. 2, left panel in \citealt{milone17}).

\section{Observed fraction of enriched stars}

From the cluster mass loss derived in the previous section, we can determine the fraction of enriched stars present in the cluster following two assumptions. Firstly, the initial amount of enriched material predicted by stellar evolution is considered to be in the range $f_{\rm en}^{\rm init}\simeq5$--$10$ per cent (see \citealt{bastian17}). The exact value depends on the type of polluters, the stellar IMF and the mass range considered. We assume it to be $f_{\rm en}^{\rm init}=5$ per cent, but augmenting the value does not produce major differences. In order to solve the mass budget problem, MBL multiple population models assume that all cluster mass loss is in the form of unenriched stars. With the aim of being conservative, we use the same assumption to determine the fraction of enriched stars in each of our surviving stellar clusters. To do that, we correct the initial-to-present mass ratio determined in the previous section for stellar evolutionary mass loss, as it affects both populations equally if they have the same IMF. We examine the dependence of the fraction of enriched stars on galactic environment and compare it to recent observations. 

Assuming both populations have the same IMF, we determine the fraction of enriched stars as
\be 
f_{\rm en} = \dfrac{f_{\rm en}^{\rm init}}{(1-f_{\rm en}^{\rm init})\left(f_{\rm *} M_{\rm cl}^{\rm init}/M_{\rm cl}\right)^{-1} + f_{\rm en}^{\rm init}},
\label{eq:fenriched}
\ee
where $M_{\rm cl}^{\rm init}$ and $M_{\rm cl}$ are the initial and present cluster masses, $f_{\rm *} = M_{\rm *}/M_{\rm *}^{\rm init}\approx 0.4$ is a factor to correct for stellar evolutionary mass loss for our adopted Chabrier IMF using $t=10~\gyr$, and $f_{\rm en}^{\rm init} = 5$ per cent is the assumed initial enriched fraction.

\begin{figure*}
\centering
\includegraphics[width=\hsize,keepaspectratio]{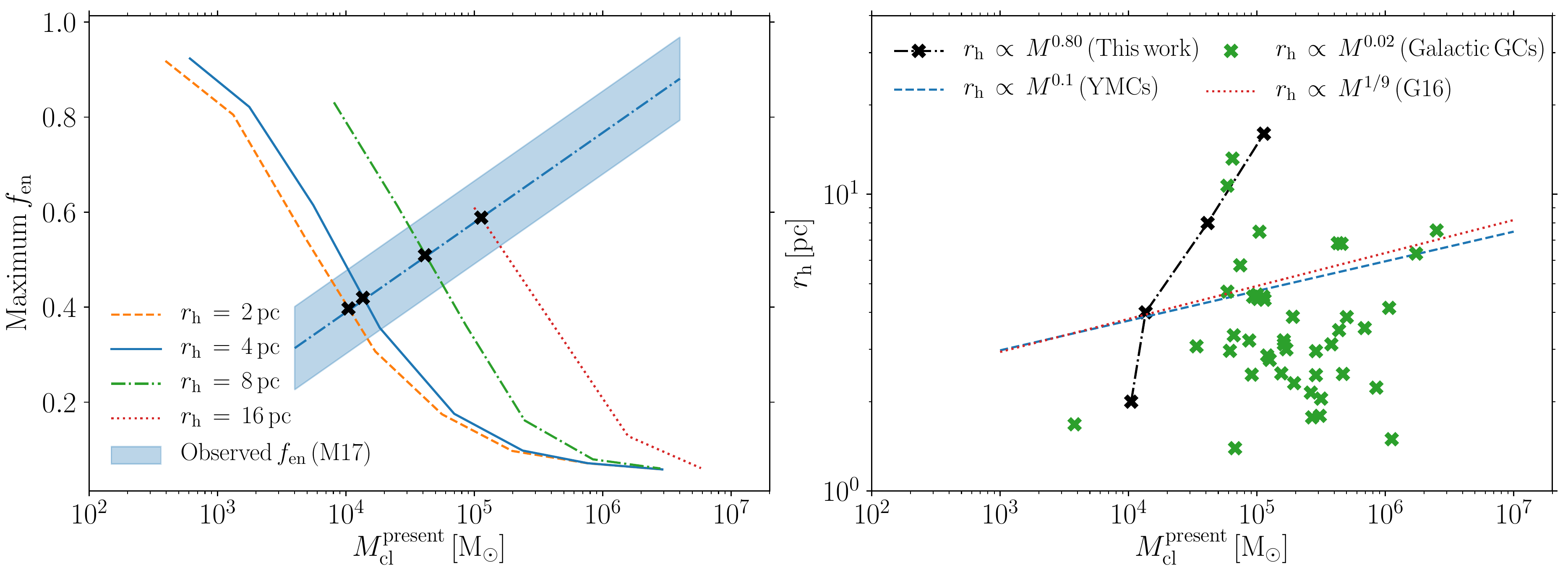}%
\vspace{-2mm}\caption{\label{fig:mass-loss-different-size} \textit{Left}: Maximum enriched fraction of old clusters born in high-pressure environments as a function of the present cluster mass for simulations with different cluster sizes, assuming an initial enriched fraction of $5$ per cent. As in Fig.~\ref{fig:fenriched-pressure}~(a), the blue dash-dotted line and area are the fiducial relation and $1\sigma$ deviation from \citet{milone17}. \textit{Right}: Half-mass radius versus present cluster mass. The black dash-dotted line is our derived mass-radius relation needed to reproduce the observed relation in the left panel through cluster mass loss, whereas the dashed blue line is the mass-radius relation found for YMCs in \citet{larsen04b} and the green crosses represent the same GCs as in the \citet{milone17} sample (\citealt{harris96}, \citealt{mclaughlin05}). The red dotted line represents the analytical mass-radius relation derived in \citet{gieles16} for the Galactic disc. All proportionalities are indicated in the legend.}\vspace{-3mm}
\end{figure*}

We present the obtained enriched fractions as a function of the cluster mass, and their medians at fixed cluster mass as a function of present galactocentric radius and metallicity in Fig.~\ref{fig:fenriched-pressure}. In the left panel we include the observed relation from \citet{milone17}. The authors determine the fraction of unenriched stars for 43 clusters and show them against the photometric cluster masses from \citet{mclaughlin05}.\footnote{Using the photometric rather than the dynamical cluster masses underestimates masses below $M_{\rm cl} < 10^4~\msun$ due to dynamical effects (e.g. \citealt{kruijssen08}). Changing to the dynamical cluster masses produces no effect in the observed trend.} The GC sample of \citet{milone17} shows no trends of the enriched fraction with galactocentric radius or metallicity (see also \citealt{kruijssen15b} and \citealt{bastian15}). The enriched fractions under the influence of dynamical mass loss obtained from \emosaics differ strongly from the observations, both in an absolute sense and in terms of its relation to the GC mass. The enriched fraction steeply declines with the present cluster mass and exhibits the expected trends with galactocentric radius and metallicity for intermediate mass ($10^3$--$10^5~\msun$) clusters. Our derived enriched fractions rule out cluster mass loss as the driving mechanism behind the observed trend; we find typical enriched fractions of $5$--$30$ per cent, whereas observations find $50$--$75$ per cent. In addition, the decreasing enriched fraction with increasing cluster mass predicted by MBL models contradicts the observed positive trend. The largest disagreement occurs at masses $M_{\rm cl}\ga10^6~\msun$. These are observed to contain $\sim80$ per cent of enriched stars, which would require them having lost $\sim 99$ per cent of their mass in the form of unenriched stars, but they lose less than a factor $2$ in mass by dynamical disruption mechanisms.

\section{Reconciling models and observations}

The lack of agreement between the modelled and observed enriched fractions suggests changes to the current models for GC formation and evolution of the origin of multiple populations are needed to reconcile models and observations. First of all, we discuss whether the consideration of other cluster disruption mechanisms might provide the sufficient mass loss to reproduce the (or lack of) observational trends. Alternatively, looking at Eq.~\ref{eq:fenriched}, there are two possible ways in which the modelled and observed fractions may be reconciled. Firstly, we consider whether the amount of mass lost in \emosaics may not be correct. The main uncertainty in the mass loss predicted by the \emosaics disruption model is posed by the assumption of a constant cluster radius. Because mass loss driven by tidal shocks depends on the cluster density, we can change the amount of mass loss experienced by modifying the assumed mass-radius relation. Secondly, the adopted initial enriched fraction from stellar nucleosynthesis might be incorrect. We can use our results to derive what the initial amount of enriched material should be to match to the observational data. We now explore these possibilities.

\subsection{Other cluster disruption mechanisms}

A large body of literature explores which mechanisms influence cluster evolution, and finds that the dominant driver of mass loss are tidal shocks with the substructure of the interstellar medium \citep[e.g.][]{gieles06,kruijssen11,miholics17,pfeffer18}. In addition, in weaker tidal fields clusters are mostly affected by two-body relaxation \citep{lamers10}, so we consider these two mechanisms, along with stellar evolution and dynamical friction, as the sources of cluster mass loss in \emosaics.

In addition to the above, other disruption mechanisms have been invoked to solve the mass budget problem, with stellar evolution-driven expansion being the preferred mechanism. \citet{dercole08} argue that only early mass loss due to the expansion associated with supernovae would lead to a strong preferential loss of the unenriched population in massive ($M=10^7~\msun$) stellar clusters. However, the efficiency of this mechanism depends on the cluster structure, IMF and the initial degree of mass segregation. Tidally-filling stellar clusters are mostly affected by stellar evolution-driven expansion mass loss if they have a high degree of mass segregation, as stellar evolutionary mass loss then lead to an enhanced flow of stars over the tidal boundary \citep{vesperini09}. The amount of mass lost is roughly inversely proportional to the galactocentric radius, with a steeper dependence for primordially mass segregated clusters \citep{haghi14}, and presents a similar dependence as the one found for evaporation-driven mass loss \citep{baumgardt03}. The lack of a self-consistent model for this mechanism in the literature prevents its implementation in \emosaics, but we can compare the relation between the expected mass loss and galactocentric radius to our numerical results. Looking at the middle panel of Fig.~\ref{fig:fenriched-pressure}, the addition of the mass lost due to stellar evolution-driven expansion would increase and steepen the enriched fractions, thus making the modelled fractions more incompatible with the constant observed fractions. Furthermore, in appendix D in \citet{pfeffer18} we test the amount of cluster disruption in our simulations and we find that compact ($r_{\rm h} = 2.2~\pc$) clusters with adiabatic expansion due to stellar evolution can double their sizes ($r_{\rm h} = 3$--$4~\pc$) in a short timescale. This implies that, under the disruption mechanisms included in \emosaics, these expanded clusters undergo the same mass loss as clusters that are equally extended at birth, suggesting that adiabatic, stellar evolution-driven expansion is indistinguishable from the regular evolution of an already extended cluster. This indicates that stellar evolution-driven expansion is not the remaining mechanism that could help explain the required mass loss to match observations.

\subsection{Mass-radius relation required by the observations}\label{subsec:mass-radius}

The fiducial model in \emosaics considers all clusters to have a constant half-mass radius of $r_{\rm h} = 4~\pc$, which means that the cluster density is solely controlled by its mass. From the dynamical disruption mechanisms considered in our model (i.e. evaporation and tidal shocking), the latter is dominant \citep[e.g.][]{miholics17,pfeffer18}. Tidal shocks are inversely proportional to the cluster's density \citep{spitzer58}, so, at fixed mass, extended clusters are expected to lose more mass than their compact counterparts.

From our sample of 25 present-day Milky Way-mass simulations, we run the halo MW05 with different cluster sizes that do not evolve with time ($r_{\rm h}\,=\,2,\,4,\,8,\,16,\,32$ and $64\,\pc$). We discard the simulations with radii of $32$ and $64~\pc$, as disruption is so efficient that in total only 56 and 3 clusters are left, respectively. We show the maximum enriched fraction (which is analogous to the upper envelope in the left-hand panel of Fig.~\ref{fig:fenriched-pressure}) of the old clusters born in a high-pressure environment as a function of the present cluster mass in Fig.~\ref{fig:mass-loss-different-size} (left). Changing the cluster radii indeed affects the enriched fraction as expected; for a given cluster mass, extended clusters lose more mass, resulting in a higher enriched fraction.

Using the modelled enriched fractions for different radii and their intersection with the observed relation between the enriched fraction and the cluster mass in Fig.~\ref{fig:mass-loss-different-size} (left), we derive the cluster mass-radius relation needed to reproduce the observed relation through cluster mass loss. We show this mass-size relation in Fig.~\ref{fig:mass-loss-different-size} (right), along with the observed mass-size relations for YMCs from \citet{larsen04b}, for the GC sample in \citet{milone17} using the cluster sizes from \citet{harris96} and the photometric cluster masses from \citet{mclaughlin05}, and the theoretical relation from \citet{gieles16}. The required mass-size relation has a much steeper slope than is supported by the observations. If we extrapolate this relation, a cluster of mass $M_{\rm cl}^{\rm present}=10^6~\msun$ should have a half-mass radius of $r_{\rm h}\sim96\,\pc$, which is not observed. 

\citet{vanzella17c} observe a candidate proto-GC of mass $2$--$4\times10^6~\msun$ and effective radius $\simeq20~\pc$ at $z=6.145$, but this radius is likely overestimated due to protoclusters forming in larger structures (e.g.~\citealt{longmore14}), that may be unresolved. Its size is an order of magnitude smaller than the half-mass radius required for cluster mass loss to lead to the high observed enriched fractions, $r_{\rm h}\sim166-289~\pc$.

The disproportionate sizes of massive clusters required to match the observed enriched fractions indicate that changing the cluster radii does not enable us to reconcile our results with the observations. This implies that cluster mass loss is not responsible for the relation between enriched fraction and cluster mass.

\subsection{Initial fraction of enriched stars}

Current multiple population models assume an initial enriched fraction of $5$--$10$ per cent (e.g. see Sect.~5.4 in \citealt{bastian17}). The exact fraction depends on the pollutor and the IMF, and assumes that all the polluted material is later on used to form enriched stars (i.e. $100$ per cent efficiency in recycling the polluted material). 

\begin{figure}
\centering
\includegraphics[width=\hsize,keepaspectratio]{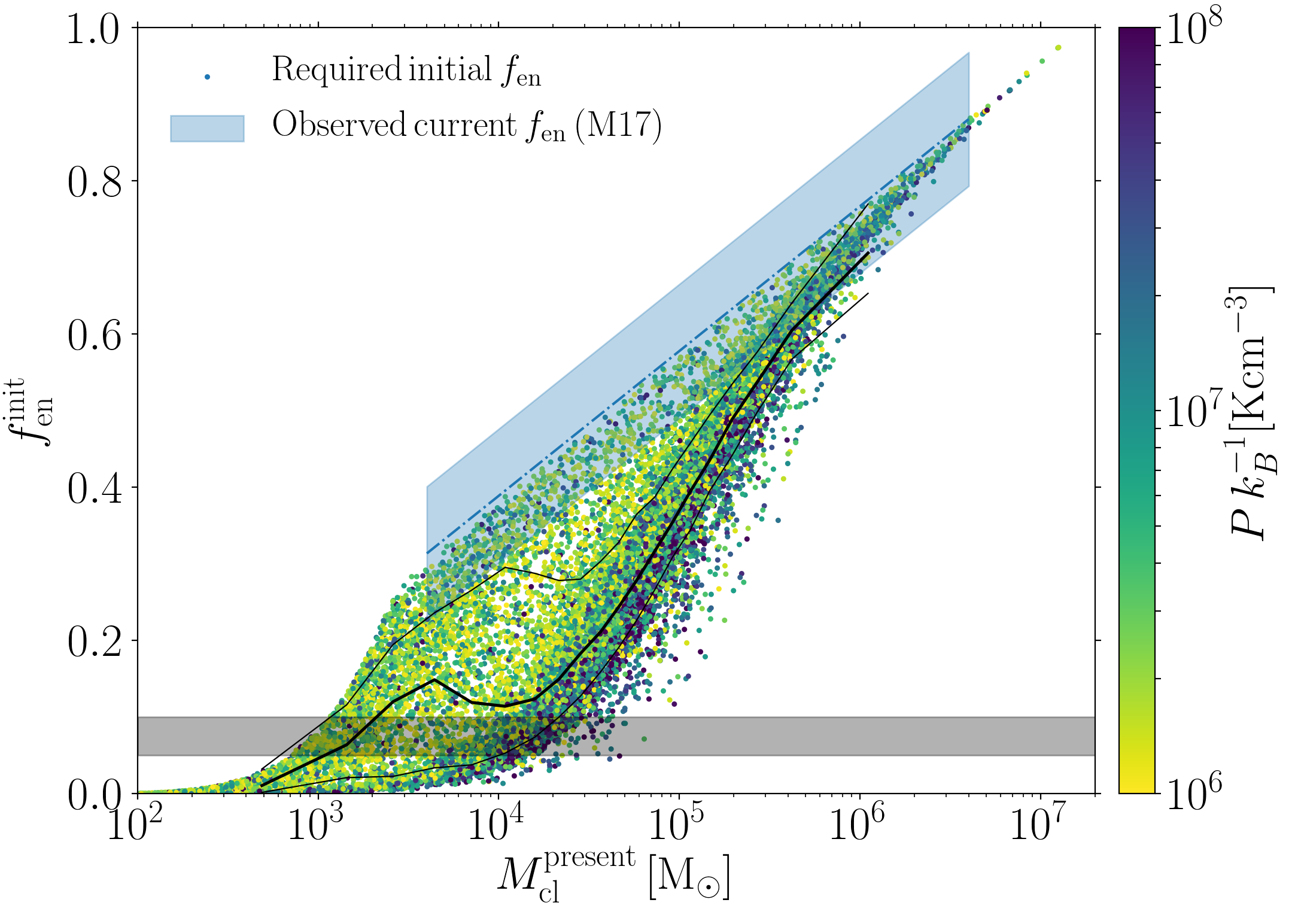}%
\vspace{-2mm}\caption{\label{fig:initial-frac-enriched} Required initial fraction of enriched stars as a function of the present cluster mass. Data points correspond to old ($\tau>6~\gyr$) clusters born in high-pressure environments ($P\,k_{\rm B}^{-1}>10^6\,\rm K\,cm^{-3}$), coloured by their gas pressure at birth. The grey area corresponds to the commonly assumed initial enriched fraction range of $5$--$10$ per cent (\citealt{bastian17}).}\vspace{-3mm}
\end{figure}

To explore whether we can reconcile the initial-to-present GC mass ratios from \emosaics with the observed enriched fraction as a function of GC mass, we invert Eq.~\ref{eq:fenriched} to determine the required initial enriched fraction given the modelled mass loss as, 
\be 
f_{\rm en}^{\rm init} = \dfrac{f_{\rm en}}{f_{\rm en}\left(1-f_{\rm *} M_{\rm cl}^{\rm init}/M_{\rm cl}\right) + f_{\rm *} M_{\rm cl}^{\rm init}/M_{\rm cl}},
\label{eq:fenriched-initial}
\ee
where $f_{\rm en}$ is the fiducial fit to the observed enriched fraction obtained from in \citet{milone17}. Fig.~\ref{fig:initial-frac-enriched} shows the required initial enriched fraction as a function of the present cluster mass. Given the cluster mass loss implied by \emosaics, the commonly-adopted intial enriched fraction of $5$--$10$ per cent is only consistent with the observed enriched fraction for clusters with masses in the range $10^3$--$10^4~\msun$. By contrast, the required initial enriched fraction for more massive clusters correlates with present cluster mass. This indicates that clusters with masses $10^5$--$10^6~\msun$ require initial enriched fractions of $f_{\rm en}^{\rm init}=0.4$--$0.7$ to reproduce the relation from \citet{milone17} after their subsequent dynamical mass loss. This range represents an order-of-magnitude departure from the commonly adopted initial enriched fractions in MBL models. 

The disparity between the commonly used range and our results indicates that either different polluters have to be considered (i.e. that produce greater amounts of ejected material per unit cluster mass, e.g. SMSs, see \citealt{gieles18}) or some yet-to-be-understood physical processes in stellar evolution and/or star formation lead to the large ($>0.4$) enriched fractions.

\section{Discussion}

We investigate whether dynamical mass loss from GCs under realistic conditions satisfies the requirements from current multiple population models for reproducing the observed fractions of enriched stars in GCs. To do so, we use all surviving clusters present in the 25 present-day Milky Way-mass galaxy simulations from the \emosaics project \citep{pfeffer18,kruijssen18b}. We determine the initial-to-present cluster mass ratios and study their dependence on cluster properties in Fig.~\ref{fig:mass-loss-pressure}. We find a steep decrease of the amount of mass lost with present cluster mass. For intermediate-mass ($10^3$--$10^5~\msun$) clusters, we also find significant trends of increasing mass loss towards smaller galactocentric radii, higher gas pressures at birth, and higher cluster metallicities. More massive clusters show no dependence between their median mass ratio and these quantities due to their small amount of mass lost.

Assuming all dynamical mass loss is in the form of unenriched stars, as posited by current multiple population models, we determine the enriched fraction of our stellar clusters. We study its dependence on cluster mass, metallicity, and galactocentric radius and compare it to observed enriched fractions from \citet{milone17} in Fig.~\ref{fig:fenriched-pressure}. These mirror the dependences found for the amount of mass loss and fall below the observed fractions by a factor of $2$--$20$, with an opposite dependence on GC mass. These severe discrepancies rule out cluster mass loss as the driving mechanism behind the high observed enriched fractions.

In order to reconcile our results with observations, we consider what can be changed in models to match the observed enriched fractions. The addition of stellar evolution-driven expansion mass loss, which is the preferred disruption mechanism suggested in the literature to solve the mass budget problem, predicts an even steeper trend between the numerical enriched fractions and galactocentric radius, in clear contrast with observations. These can only be reproduced if we consider an unphysically steep cluster mass-size relation, which indicates dynamical disruption mechanisms cannot account for the amount of mass loss required by current multiple population models. We determine the initial amount of polluted material required to match our results with the observations. The commonly adopted range of $f_{\rm en}=5$--$10$ per cent is only valid for clusters with masses $M_{\rm cl}^{\rm present} \lesssim 10^3$--$10^4~\msun$. At higher masses, initial enriched fractions of $f_{\rm en}=10$--$80$~per~cent are required. We conclude that dynamical cluster disruption mechanisms are not capable of explaining the high enriched fractions observed in GCs nor their positive trend with cluster mass. Hence, the present-day enriched fractions likely reflect their initial values, fundamentally challenging most self-enrichment scenarios which are incapable of producing fractions above $10$--$20$ per cent.

\vspace{-2mm}\section*{Acknowledgements}
We thank the anonymous referee for their helpful comments and suggestions. MRC is supported by a PhD Fellowship from the IMPRS-HD. MRC and JMDK acknowledge funding from the European Research Council (ERC-StG-714907, MUSTANG). JMDK acknowledges funding from the German Research Foundation (DFG - Emmy Noether Research Group KR4801/1-1). NB and JP gratefully acknowledge financial support from the European Research Council (ERC-CoG-646928, Multi-Pop). This work used the DiRAC Data Centric system at Durham University, operated by the Institute for Computational Cosmology on behalf of the STFC DiRAC HPC Facility. This equipment was funded by BIS National E-infrastructure capital grant ST/K00042X/1, STFC capital grants ST/H008519/1 and ST/K00087X/1, STFC DiRAC Operations grant ST/K003267/1 and Durham University. DiRAC is part of the National E-Infrastructure. The work also made use of high performance computing facilities at Liverpool John Moores University, partly funded by the Royal Society and LJMU’s Faculty of Engineering and Technology.

\bibliographystyle{mnras}
\vspace{-2mm}\bibliography{bibdesk-bib}

\bsp

\label{lastpage}

\end{document}